 \documentclass[preprint,aip,draft, double-spaced]{revtex4-1}

\usepackage{color}
\usepackage{graphicx}
\usepackage[english]{babel}
\usepackage{amsfonts,amssymb,amsmath}

\usepackage[latin1]{inputenc}

\begin{document}

\preprint{APS/123-QED}

\title{Room Temperature Coherent and Voltage Tunable Terahertz Emission from Nanometer-Sized Field Effect Transistors}


\author{S. Boubanga-Tombet}
\affiliation{Research Institute of Electrical Communication, Tohoku University, 2-1-1 Katahira, Aoba-Ku, Sendai 980-8577, Japan}
\author{F. Teppe\email{frederic.teppe@ges.univ-montp2.fr}}
\affiliation{Groupe d'\'Etude des Semiconducteurs, UMR 5650 CNRS, Universit\'e Montpellier 2, 34095 Montpellier, France}
\author{J. Torres\email{jeremi.torres@ies.univ-montp2.fr}} 
\affiliation{Institut d'\'Electronique du Sud, UMR 5214 CNRS, Universit\'e Montpellier 2, 34095 Montpellier, France}
\author{A. El Moutaouakil}
\affiliation{Research Institute of Electrical Communication, Tohoku University, 2-1-1 Katahira, Aoba-Ku, Sendai 980-8577, Japan}
\author{D. Coquillat, N. Dyakonova, C. Consejo, P. Arcade}
\affiliation{Groupe d'\'Etude des Semiconducteurs, UMR 5650 CNRS, Universit\'e Montpellier 2, 34095 Montpellier, France}
\author{P. Nouvel, H. Marinchio, T. Laurent, C. Palermo, A. Penarier} 
\affiliation{Institut d'\'Electronique du Sud, UMR 5214 CNRS, Universit\'e Montpellier 2, 34095 Montpellier, France}
\author{T. Otsuji}
\affiliation{Research Institute of Electrical Communication, Tohoku University, 2-1-1 Katahira, Aoba-Ku, Sendai 980-8577, Japan}
\author{L. Varani} 
\affiliation{Institut d'\'Electronique du Sud, UMR 5214 CNRS, Universit\'e Montpellier 2, 34095 Montpellier, France}
\author{W. Knap}
\affiliation{Groupe d'\'Etude des Semiconducteurs, UMR 5650 CNRS, Universit\'e Montpellier 2, 34095 Montpellier, France}

\begin{abstract}
We report on reflective electro-optic sampling measurements of TeraHertz emission from nanometer-gate-length InGaAs-based high electron mobility transistors. The room temperature coherent gate-voltage tunable emission is demonstrated. We establish that the physical mechanism of the coherent TeraHertz emission is related to the plasma waves driven by simultaneous current and optical excitation. A significant shift of the plasma frequency and the narrowing of the emission with increasing channel's current are observed and explained as due to the increase of the carriers density and drift velocity.

 \end{abstract}

\maketitle

Nanometer sized transistors are extremely interesting for generation and detection of TeraHertz (THz) radiations because of the possibility of their integration with other opto-electronic devices. Plasma frequencies in nano-transistors channels'  fall effectively in the THz range and are expected to be tunable by the gate voltage \cite{Chaplik:1972vn,Dyakonov:1993sj}. The emission of THz radiations has been observed in different field effect transistors (FETs) structures \cite{Dyakonova2005,dyakonova:141906,Hirakawa1995,Hopfel1982,knap:2331,Otsuji2006,Tsui1980}. One of the mechanisms responsible for tunable THz emission by FETs was proposed by Dyakonov and Shur \cite{Dyakonov:1993sj}. In the frame of this theory, the electron flow in the channel can be unstable because of plasma waves amplification by reflections on the channels' borders.
The first observed THz generation from nano-transistors -- which has been interpreted trough the Dyakonov-Shur instability -- appears with a threshold-like behavior at a certain drain bias \cite{knap:2331}. The measured emission frequency corresponded to the lowest fundamental plasma mode in the gated region of the device. Later, Otsuji et al.\cite{OtsujiJPhys2008} have proposed a novel high electron mobility transistor (HEMT) structure based on GaInAs with doubly interdigitated grating gates, succeeding in room temperature generation of THz radiations due to the coherent plasmon excitation\cite{meziani:201108,ElMoutaouakil2009}. With standard GaN HEMT structures, El Fatimy et al.\cite{ElFatimy-el-lett-2006} have reported on incoherent room temperature THz emission which was gate-voltage tunable.  

In this letter we report on the first room temperature coherent and tunable THz emission from in InGaAs-based HEMTs. A current driven spectral narrowing is also reported and interpreted as due to the enhancement of the carriers drift velocity.

\begin{figure}[htbp]
\includegraphics[width=0.9\columnwidth]{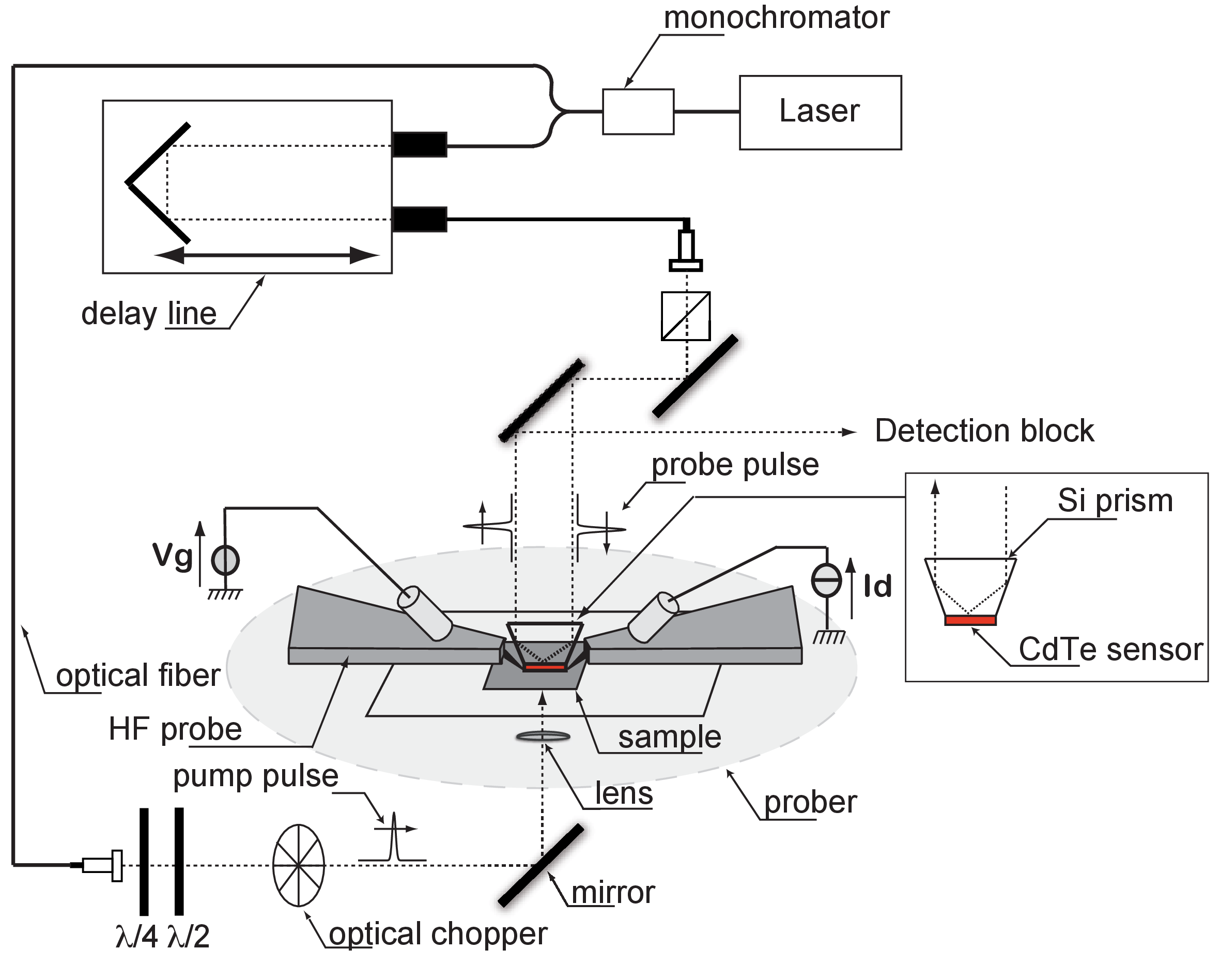}
\caption{\label{setup}Experimental configuration of electro-optic sampling for measurements of THz emission.}%
\end{figure}

Experiments were performed on GaInAs HEMTs from InP technology with a gate-length value $L_g$ = 200 nm. The threshold voltage was $V_{th} \approx -300$\;mV. More precise description of HEMTs layers can be found in Ref.\cite{teppe:222109}. 
The room temperature THz emission measurements were performed using a reflective electro-optic sampling (REOS) system (Fig. \ref{setup}). This time-resolved spectroscopy technique uses femtosecond pump and probe 1.55-$\mu$m laser beams with 80 fs pulse duration and 20 MHz repetition rate\cite{Otsuji2006}. The linearly polarized pump beam illuminating the device from the rear surface creates photocarriers via resonant interband excitation. The femtosecond profile of the photogenerated carriers can be decomposed in a sum of harmonics that excites all THz plasma modes. Therefore, the optical excitation induce THz generation by the enhancement of the plasma waves amplification. 

\begin{figure}[htbp]
\includegraphics[width=0.85\columnwidth]{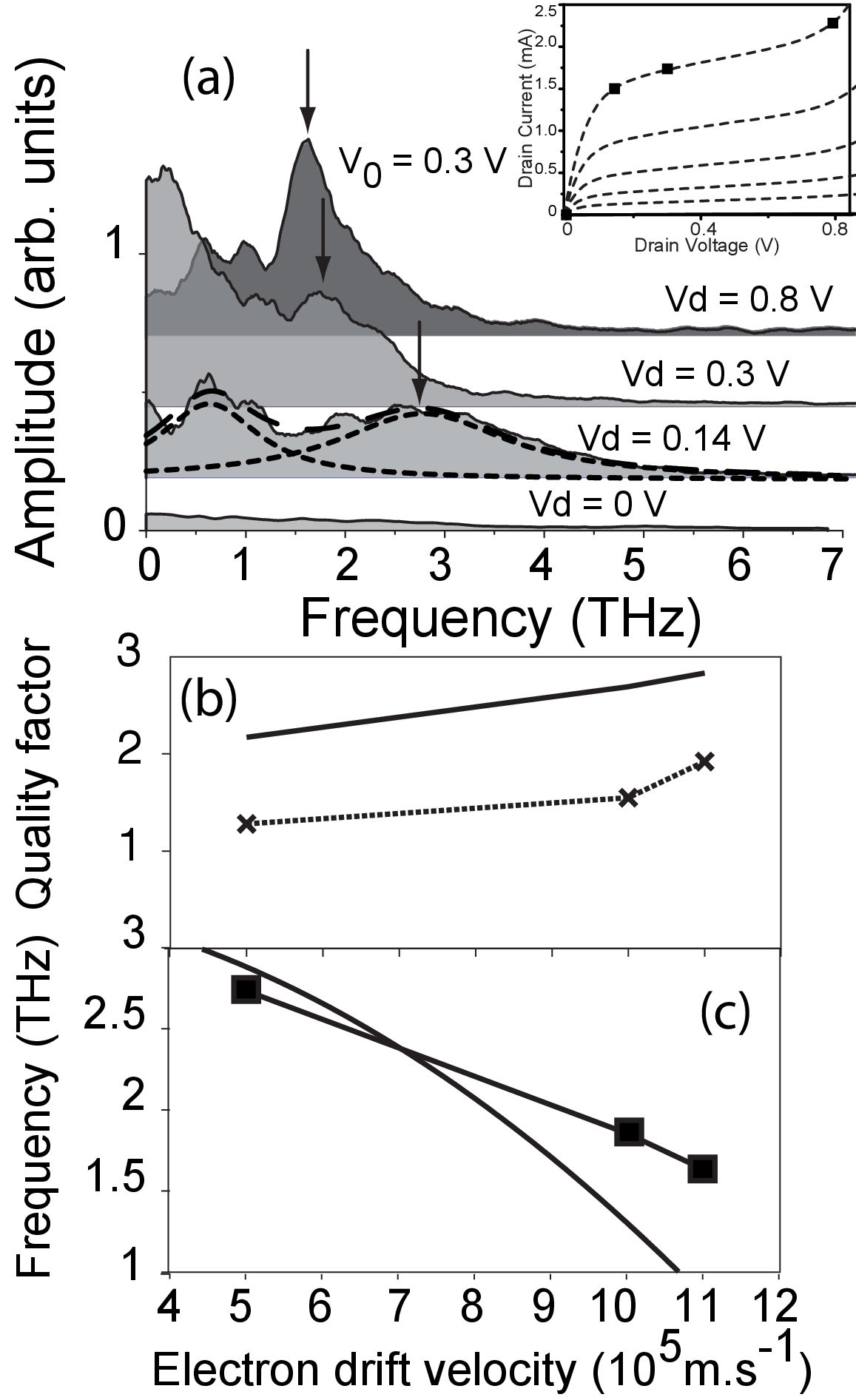}
\caption{\label{deltaU_vd} (a) Emission spectra by varying the drain voltage from 0 V up to 0.8 V. Dashed and dotted lines are Lorentzian fits of curve at $V_0$ = 0.3 V and $V_d$ = 0.14 V. Insert:  static output characteristics for a gate potential varying from -0.2 V to 0 V with 0.05 V step under illumination. (b) Quality factor and (c) peak frequency variation as a function of the average electron drift velocity .}%
\end{figure}

Concerning the detection part of the set-up, a 120 $\mu$m thick-CdTe electro-optic sensor crystal with Si prism was put on the sample top surface. The probe beam, roughly cross polarized with respect to the pump beam, was led to the sensor normal to the prism surface, which is fully reflected at the CdTe surface back to the electro-optic detection block. The reflective index tensor is modulated proportionally to the electric field arisen at the crystal surface by Pockels effect. The electric field is then detected as polarization change of the probe beam. Its temporal profile is reconstructed as a function of the delay between pump and probe pulse. The Fourier transform of the obtained temporal profile gives the THz spectra emitted by the sample. As the pump induced THz field is perfectly correlated with itself over the repetition rate, this stroboscopic technique gives access to the coherent part of the emitted field. To fulfill the plasma theory conditions, the HEMTs where :  (i) connected in common source configuration and (ii) biased at the drain by an external current source\footnote{In the following, the applied drain current values are given in terms of their corresponding drain-voltage values for readability consideration with the insert of Fig. \ref{deltaU_vd}}.

\begin{figure}[htbp]
\includegraphics[width=0.85\columnwidth]{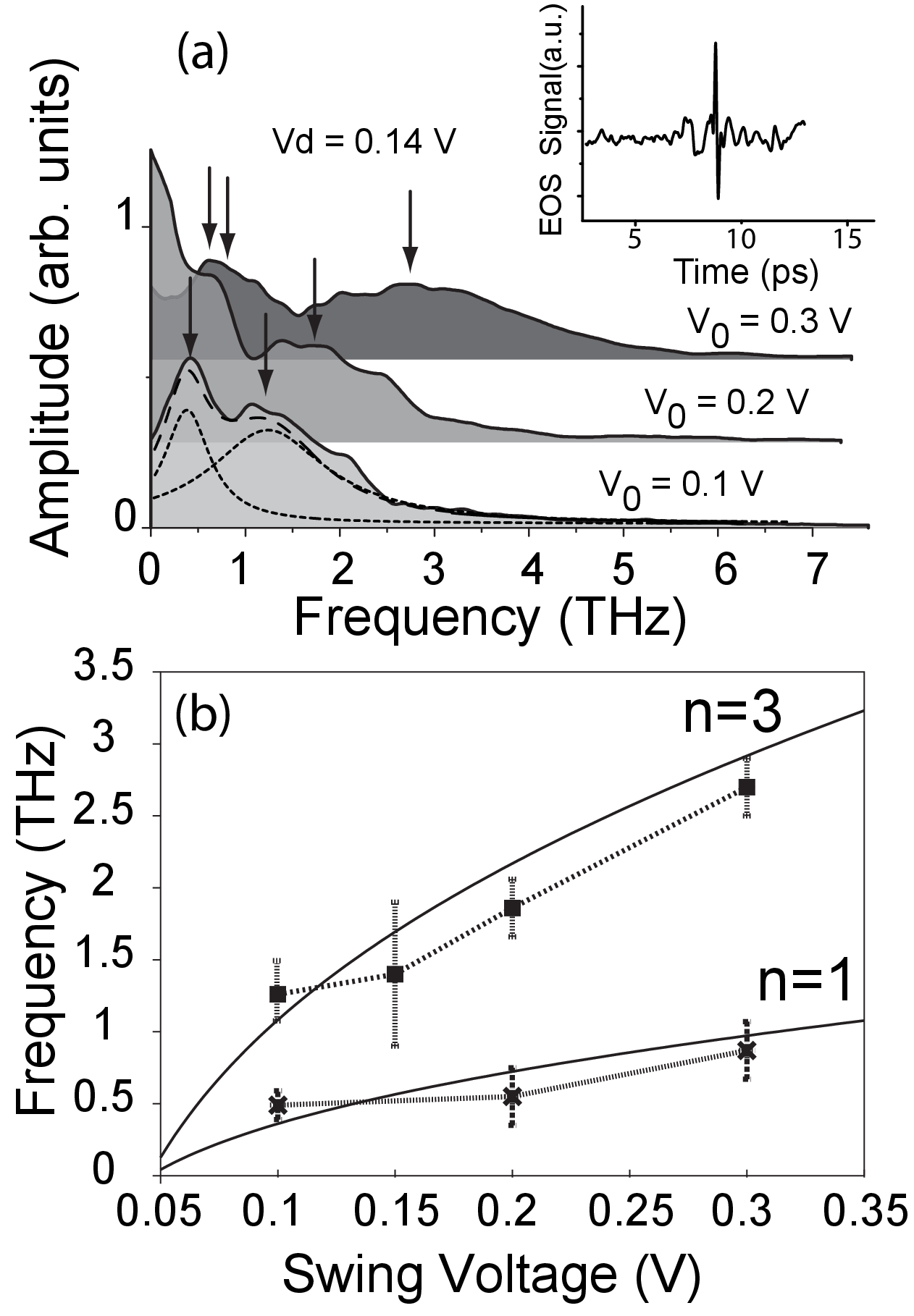}%
\caption{\label{deltaU_vg} (a) Measured field emission spectrum when the pump beam is shinning the device varying the swing voltage from 0.1 V up to 0.3 V for $V_d$=0.14 V. Dashed and dotted lines are Lorentzian fits at $V_0$ = 0.1 V. Insert: temporal response of the emission spectra. (b) Frequency dependence of the peaks with the swing-voltage. Solid lines are calculated using eq (\ref{eq:un}) with n=1 and n=3.}%
\end{figure}

Figure \ref{deltaU_vg} (a) shows an example of measured interferogram (insert) and the smoothed Fourier transformed spectra for different swing voltages from $V_0 = V_g - V_{th}$ = 0.1 V to $V_0$ = 0.3 V at constant drain voltage of $V_d$ = 0.14 V. Lorentzian fits of the emission spectra have been used to determine the peak positions. A double Lorentzian fit (dotted and dashed lines) is also presented on the curve at $V_0$ = 0.1 V to show that the wide-band coherent emission bump is composed by two peaks. Both peaks shift to higher frequency values with the swing voltage increasing from 0.1 V up to 0.3 V. The frequency dependence of these peaks as a function of the swing voltage is reported in figure \ref{deltaU_vg} (b). The low frequency peak (n=1, crosses) varies from 0.49 THz to 0.87 THz whereas the high frequency peak (n=3, squares) varies from 1.26 THz to 2.7 THz. The solid lines are calculated using \cite{Dyakonov:1993sj}:
\begin{equation}\left \{ \begin{array}{ll}
\displaystyle{ f_n = \frac{n}{4}  \frac{s}{L_{eff}} (1-\frac{v^2}{s^2})}\\
 \displaystyle{ n = 1, 3, 5\dots}
  \label{eq:un}
\end{array}\right. \end{equation}
where $f_n$ is the frequency of the $n^{th}$ plasma wave mode, $s = \sqrt{(eV_0/m^{*})} $ is the plasma wave velocity, $e$ is the electron charge, $m*$ is the electron effective mass in $\textrm{In}_{0.53}\textrm{Ga}_{0.47}\textrm{As}$ and $v$ = 5.10$^5$ m.s$^{-1}$ is the  average electron drift velocity estimated using an hydrodynamic model \cite{marinchio:192109} at $V_0$ = 0.3 V and $V_d$ = 0.14 V. We assume that the effective gate length is the length of channel controlled by the gate and defined as in ref. \cite{ElFatimy-el-lett-2006,knap:2331} by $L_{eff} = L_g + 2d$ = 240 nm, where $d$ is the gate-to-channel distance. As one can note, the position of the emission peaks as a function of the gate bias follows the theoretical prediction. These results allow us to attribute the emission spectra to the optically enhanced plasma wave resonances. The lower frequency curve (n=1) corresponds to fundamental plasma mode and the higher one (n=3) to the third harmonic.

The insert of Fig. \ref{deltaU_vd} (a) shows the static output characteristics taken under illumination for a gate potential varying from - 0.2 V to 0 V with 0.05 V step. Squares indicate the applied drain voltages used in Fig. \ref{deltaU_vd} (a). The latter shows the variation of the emitted field spectrum as a function of the applied drain voltage for swing voltage kept constant at $V_{0}$ = 0.3 V. The first curve corresponds to $V_d$ = 0 V, for which a very weak emission is measured. When the drain voltage reaches 0.14 V, a double peak appears  (see also Fig. \ref{deltaU_vg}). These results show that application of the drain current is necessary to trigger the THz generation. With increasing the drain current (i.e. the electron drift velocity), the amplitude of the high frequency peak increases and its linewidth clearly shrinks. A shift of this peak position to lower frequencies is also observed. In a FET with a dc current flowing in the channel, the plasma waves are carried along the channel by the electron flow \cite{Dyako-IEEE-96}. Therefore the drain current acts against the plasma wave damping and thus, induces a shrinking of the resonance linewidth. 

According to ref. \cite{boubanga-tombet:212101} the quality factor in the current driven regime is given by :
 \begin{equation}
Q=\omega \tau = 2 \pi f \frac{\tau L_{eff}}{L_{eff} - 2v\tau}
\label{eq:deux}
\end{equation}  
where $\tau$ is the momentum relaxation time. Fig. \ref{deltaU_vd} (b), shows the quality factor of the resonances linewidth experimentally determined by $Q = f/\Delta f$ as a function of the average electron drift velocity. This velocity was estimated using hydrodynamic model of ref. \cite{marinchio:192109} with our sample parameters. The crosses are experimental results and the solid line is calculated using Eq. (\ref{eq:deux}). Figure \ref{deltaU_vd} (c) illustrates the peak frequency variation. The theoretical solid line was obtained by taking the same values of velocity. The squares are experimental results whereas the solid line is a calculation using Eq. (\ref{eq:un}). For both physical quantities, experimental results and calculations are in good quantitative agreement with plasma wave emission theory.

In conclusion, we have shown that the room temperature emission of Terahertz electromagnetic waves from InGaAs nanometric HEMTs is coherent and can be tunable with the gate voltage. We have also shown that increasing the drain bias significantly modifies the emission spectrum by a red-shift and a clear narrowing of the resonance lines. The physical mechanism of the emission was shown to be due to the plasma wave resonances optically excited in the transistor channel under current driven regime. We have demonstrate that properly excited nanotransistors can pave the way for a new class of coherent, room temperature, and easily tunable THz sources.  

\section*{Acknowledgments}
Authors wish to thank Pr. M. Dyakonov for valuable discussions, Pr. S. Bollaert and Dr. Y. Rollens (IEMN-UMR 5820) for providing the InGaAs transistors. This work was done in the frame of the TERALAB - GIS group. Supports from the SAKURA French-Japan grant, from the Japan Grant in Aid for Scientific Research from JSPS, from GDR-I Semiconductors sources and detectors of Terahertz radiation network and from the JST-ANR WITH project are also acknowledged.

\newpage

%

\newpage

\end{document}